\documentclass[aps,prb,superscriptaddress,showpacs,amssymb,secnumarabic,twocolumn,floatfix]{revtex4}

\usepackage{graphicx}
\usepackage[latin1]{inputenc}
\usepackage{dcolumn}
\usepackage{bm}
\usepackage{amssymb}
 \usepackage{hyperref}

\def\va1{\vec{a}_{1}}

\def\vb1{\vec{b}_{1}}

\def\vd1{\vec{\delta}_{1}}

\newcommand{\ba}{\begin{eqnarray}}
\newcommand{\ea}{\end{eqnarray}}
\def\be{\begin{equation}}
\def\ee{\end{equation}}

\def\vk{\vec{k}}
\def\vx{\vec{x}}
\def\vy{\vec{y}}

\def\vra{\vec{r}_{\alpha}}
\def\vrb{\vec{r}_{\beta}}

\def\vk{\mathbf{k}}

\newcommand{\op}[1]{ \hat{#1}}
\newcommand{\ve}[1]{ \mathbf{#1}}

\begin{document}


\title{Evidence of a spin liquid phase in the frustrated honeycomb lattice}

\author{D.\ C.\ Cabra}
\affiliation{Instituto de F\'\i sica de La Plata and Departamento de F\'isica, Universidad Nacional de La Plata, C.C. 67, 1900 La Plata, Argentina}
\affiliation{Facultad de Ingenier\'\i a, Universidad Nacional de Lomas de Zamora, Cno.\ de Cintura y Juan XXIII, (1832) Lomas de Zamora, Argentina.}
\author{C.\ A.\ Lamas}
\affiliation{Instituto de F\'\i sica de La Plata and Departamento de F\'isica,
 Universidad Nacional de La Plata, C.C. 67, 1900 La Plata, Argentina}
\author{H.\ D.\ Rosales}
\affiliation{Instituto de F\'\i sica de La Plata and Departamento de F\'isica,
 Universidad Nacional de La Plata, C.C. 67, 1900 La Plata, Argentina}

\begin{abstract}
In the present paper we present some new data supporting the existence of a spin-disordered
phase in the Heisenberg model on the honeycomb lattice with antiferromagnetic interactions up to third neighbors
along the line $J_2=J_3$, predicted in \cite{nos}. We use the Schwinger boson technique followed by a mean field decoupling and
exact diagonalization for small systems to show the existence of an intermediate phase with a spin gap and
short range N\'eel correlations in the strong quantum limit $(S=\frac12)$.
\end{abstract}

\maketitle


\section{Introduction}

Geometrical frustration in two-dimensional
(2D) antiferromagnets is expected to enhance the effect of quantum spin fluctuations and hence suppress
magnetic order giving rise to a spin liquid \cite{Anderson} and this idea has motivated many researchers
to look for its realization \cite{Sachdev1,Sachdev2,Sandvik,Moessner,Poilblanc}.

One candidate to test these ideas is the honeycomb lattice, which is bipartite
and has a classical N\'eel ground state, but due to the small coordination number $(z=3)$,
quantum fluctuations could be expected to be stronger than those in the square
lattice and may destroy the antiferromagnetic long-range order (LRO) \cite{Mattsson,more}.

The study of frustrated quantum magnets on the honeycomb lattice
has also experimental motivations \cite{ESR,libro_experimental,exp_hex2,expnew1,expnew2}.

The analysis of the hexagonal lattice from a more general point of view has gained lately a lot of interest both
coming from graphene-related issues and from the possible spin-liquid phase found in the Hubbard model in such
geometry \cite{HubbardNature,recent,newer}.

Motivated by the previous results, in this paper we show the study of the Heisenberg model  on the honeycomb
lattice with first ($J_1$), second ($J_2$) and third ($J_3$) neighbors couplings \cite{Fouet}, along the
special line $J_{2}=J_3$. Using Schwinger boson mean field theory (SBMFT) and exact
diagonalization we find strong evidence for the existence of an intermediate disordered region where a
spin gap opens and spin-spin correlations decay exponentially.
Using exact diagonalization of small clusters we also have calculated the dimer-dimer correlation function
that indicates short range dimer-dimer order. Although our results correspond to a specific line, we conjecture
that the quantum disordered phase that we have found in the vicinity of the tricritical point extends within a
finite region around it.  Previous evidence of massive behaviour in the hexagonal lattice Heisenberg model has
been found in other regions of the phase space by means of exact diagonalization in ref. \cite{Fouet}.

\begin{figure}[t]
\includegraphics[width=0.3\textwidth]{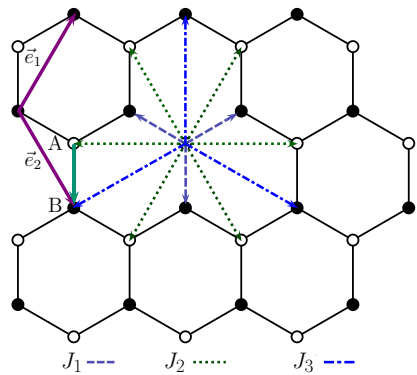}
\caption{(Color online) Honeycomb lattice with sublattices A and B. The vectors
$\vec{e}_1=(\frac{\sqrt{3}}{2},\frac{3}{2})$ and $\vec{e}_2=(\frac{\sqrt{3}}{2},-\frac{3}{2})$ are the primitive traslation
vectors of the direct lattice.}
\label{fig:red}
\end{figure}

The Heisenberg model on the $J_1-J_2-J_3$ honeycomb lattice is given by
\small
\ba
 \label{eq:Hspin_general}
 H =J_1\sum_{\langle \ve{x}\ve{y}\rangle_1} \hat{\bf{S}}_{\ve{x}}\cdot
\hat{\bf{S}}_{\ve{y}} +J_2\sum_{\langle \ve{x}\ve{y}\rangle_{2}} \hat{\bf{S}}_{\ve{x}}\cdot
\hat{\bf{S}}_{\ve{y}}+J_3\sum_{\langle \ve{x}\ve{y}\rangle_{3}} \hat{\bf{S}}_{\ve{x}}\cdot
\hat{\bf{S}}_{\ve{y}},
\ea
\normalsize
where $\hat{\bf{S}}_{\ve{x}}$ is the spin operator on site $\ve{x}$ and $\langle \ve{x}\ve{y}\rangle_n$ indicates sum over the $n$-th neighbors (see Fig.~\ref{fig:red}).
In the classical limit, $S\to\infty$, the model displays different zero temperature phases with a tricritical point at $J_2=J_3=\frac12 J_1$. At this particular point the  classical ground state has a
large GS degeneracy \cite{Rastelli,Fouet}.
The Heisenberg model on the honeycomb lattice was studied using SBMFT by
Mattsson et al \cite{Mattsson} for antiferromagnetic interactions at first
and second neighbors. Here we study the Hamiltonian (\ref{eq:Hspin_general})
using a rotationally invariant version of this technique, which has proven
successful in incorporating quantum fluctuations \cite{Trumper1,Trumper2,Coleman}.
\section{Schwinger bosons mean-field theory.}
\label{chap:SB}

In this section we describe in detail the Schwinger boson  mean field
theory used in the present work. The $SU(2)$ Heisenberg Hamiltonian on a general lattice can be written as
\ba \label{eq:H_gral_sb} \op{H}=\frac12
\sum_{\ve{x}\ve{y}\alpha\beta} J_{\alpha
\beta}(\ve{x}-\ve{y})\op{\ve{S}}_{\ve{x}+\ve{r}_{\alpha}} \cdot
\op{\ve{S}}_{\ve{y}+\ve{r}_{\beta}},
\ea
where $\ve{x}$ and $\ve{y}$ are the positions of the unit cells and vectors
$\ve{r}_{\alpha}$ correspond to the positions of  each atom within the unit cell.
$J_{\alpha \beta}(\ve{x}-\ve{y})$ is the exchange interaction between the spins located in
 $\ve{x}+\ve{r}_{\alpha}$ e $\ve{y}+\ve{r}_{\beta}$.

In what follows we assume that the classical order can be parameterized as
\ba
\op{S}^{x}_{\ve{x}+\ve{r}_{\alpha}}&=&S\sin\varphi_{\alpha}(\ve{x})\\
\op{S}^{y}_{\ve{x}+\ve{r}_{\alpha}}&=&0\\
\op{S}^{z}_{\ve{x}+\ve{r}_{\alpha}}&=&S\cos\varphi_{\alpha}(\ve{x}),
\ea
with $\varphi_{\alpha}(\ve{x})={\bf Q}\cdot\ve{x}+\theta_{\alpha}$, where  ${\bf Q}$ is the ordering vector and
  $\theta_{\alpha}$ are the relative angles between the classical spins inside each unit cell.


The spin operators  $\op{\ve{S}}_\ve{x}$ on site $\ve{x}$ are represented by two bosons  $\op{b}_{\ve{x}\sigma}$
($\sigma=\uparrow,\downarrow$)
\ba \op{\ve{S}}_\ve{r}=\frac12\; \op{\ve{b}}_\ve{r}^{\dag} \cdot
\vec{\sigma}\cdot \op{\ve{b}}_\ve{r}\, , \quad \op{\ve{b}}_\ve{r}=\left(\begin{array}{c}
  \op{b}_{\ve{r}\uparrow} \\
    \op{b}_{\ve{r}\downarrow}
     \end{array}\right),
\ea
where $\vec{\sigma}=(\sigma_x,\sigma_y,\sigma_z)$ are the Pauli matrices.
This is a faithful representation of the algebra SU(2) if we take into account the following local
constraint
\ba
\label{eq:sb constraint}
2S&=&\op{b}^{\dag}_{{\bf x}\,\uparrow} \op{b}_{{\bf x}\,\uparrow}+
\op{b}^{\dag}_{{\bf x}\,\downarrow}\op{b}_{{\bf x}\,\downarrow}.
 \ea
In this representation, the exchange term can be expressed as

\ba
 \op{\ve{S}}_{\ve{x}+\ve{r}_{\alpha}}\cdot
\op{\ve{S}}_{\ve{y}+\ve{r}_{\beta}}&=&:\op{B}^{\dag}_{\alpha\beta}(\ve{x},\ve{y})
 \op{B}_{\alpha\beta}(\ve{x},\ve{y}):-
\op{A}^{\dag}_{\alpha\beta}(\ve{x},\ve{y})
\op{A}_{\alpha\beta}(\ve{x},\ve{y}),\nonumber\\
\ea
where $\op{A}_{\alpha,\beta}(\ve{x},\ve{y})$ and
$\op{B}_{\alpha,\beta}(\ve{x},\ve{y})$ are the $SU(2)$ invariants
defined as
\ba
\label{eq:A}
 \op{A}_{\alpha,\beta}(\ve{x},\ve{y})&=&\frac12 \sum_{\sigma}
\sigma \op{b}^{(\alpha)}_{\ve{x},\sigma}
\op{b}^{(\beta)}_{\ve{y},-\sigma}\\
\label{eq:B}
\op{B}_{\alpha,\beta}(\ve{x},\ve{y})&=&\frac12 \sum_{\sigma}
\op{b}^{\dag(\alpha)}_{\ve{x},\sigma}
\op{b}^{(\beta)}_{\ve{y},\sigma},
\ea
with $\sigma=\uparrow, \downarrow$ and double dots ($:\op{O}:$) indicate normal ordering of operator $\op{O}$.
This decoupling is particularly useful to the description of magnetic systems near disordered states,
because it allows to treat antiferromagnetism and ferromagnetism in equal footing.

To construct a mean field theory we perform a Hartree-Fock decoupling
\small
\ba
 \nonumber (\op{\ve{S}}_{\ve{x}+\ve{r}_{\alpha}}\cdot
\op{\ve{S}}_{\ve{y}+\ve{r}_{\beta}})_{MF}&=& [B_{\alpha
\beta}^{*}(\ve{x}-\ve{y}) \op{B}_{\alpha \beta}(\ve{x},\ve{y})\\\nonumber
&-&
A_{\alpha \beta}^{*}(\ve{x}-\ve{y}) \op{A}_{\alpha \beta}(\ve{x},\ve{y})+H.c]\\
&-& \langle (\op{\ve{S}}_{\ve{x}+\ve{r}_{\alpha}}\cdot
\op{\ve{S}}_{\ve{y}+\ve{r}_{\beta}})_{MF} \rangle,
 \ea
\normalsize
with
\small
 \ba
 \label{eq:MF_eq_SB1}
 A_{\alpha \beta}^{*}(\ve{x}-\ve{y})&=&\langle  \op{A}_{\alpha \beta}^{\dag}(\ve{x},\ve{y})\rangle\\
 \label{eq:MF_eq_SB2}
 B_{\alpha \beta}^{*}(\ve{x}-\ve{y})&=&\langle  \op{B}_{\alpha \beta}^{\dag}(\ve{x},\ve{y})\rangle\\\nonumber
 \label{eq:MF_eq_SB3}
\langle (\op{\ve{S}}_{\vx+\vra}\!\!\cdot \!\! \op{\ve{S}}_{\vy+\vrb} )_{MF} \rangle
\! &=&\! |B_{\alpha \beta}(\vx-\vy)|^{2}-|A_{\alpha
\beta}(\vx-\vy)|^{2},
 \ea
\normalsize
and where $\langle\;\rangle$ denotes the expectation value in the ground state at $T=0$.
%
%
%
Because several functions involved in the Hamiltonian
 depend on the difference $\ve{x}-\ve{y}$  we change variables to $\ve{R}=\ve{x}-\ve{y}$
and eliminating  $\ve{x}$ in the sums we obtain
\begin{widetext}
\small
\ba
\op{H}_{MF}= \frac12 \sum_{\ve{R}\ve{y}\alpha\beta} J_{\alpha \beta}(\ve{R})
\left\{\frac12 \sum_{\sigma}\left[
B_{\alpha,\beta}(\ve{R})\;\op{b}^{\dag(\alpha)}_{\ve{R}+\ve{y},\sigma}
\op{b}^{(\beta)}_{\ve{y},\sigma}-\sigma
A_{\alpha,\beta}(\ve{R})\;\op{b}^{\dag
(\alpha)}_{\ve{R}+\ve{y},\sigma} \op{b}^{\dag
(\beta)}_{\ve{y},-\sigma} +H.C.\right] -
\left(\;|B_{\alpha,\beta}(\ve{R})|^{2}-|A_{\alpha,\beta}(\ve{R})|^{2}\right)\right\}.
\ea
\end{widetext}
\normalsize

The mean field Hamiltonian is quadratic in the boson operators and can be diagonalized in real space.
However, as we look for translational invariant solutions,
it is convenient to transform the operators to momentum space
\ba
 \op{b}^{(\alpha)}_{\ve{x},\sigma}=\frac{1}{\sqrt{N_{c}}}
\sum_{\ve{k}}\op{b}^{(\alpha)}_{\ve{k},\sigma}e^{i\ve{k} \cdot
(\ve{x}+\ve{r}_{\alpha})} .
\ea
After some algebra and using the symmetry properties:
 \ba
\label{eq:simetrias1}
\nonumber  J_{\alpha \beta}(\ve{R})&=&J_{\beta \alpha }(-\ve{R})\\
  A_{\alpha \beta}(\ve{R})&=&-A_{\beta \alpha }(-\ve{R})\\
\nonumber  B_{\alpha \beta}(\ve{R})&=&B^{*}_{\beta \alpha }(-\ve{R}),
 \ea
we obtain the following form for the Hamiltonian
\begin{widetext}
\small
 \ba
\nonumber
 \op{H}_{MF}&=&\frac12
 \sum_{\ve{k}\alpha\beta}\sum_{\sigma}\left\{ \gamma^{B}_{\alpha\beta}(\ve{k})\op{b}^{\dag(\alpha)}_{\ve{k}\sigma}
 \op{b}^{(\beta)}_{\ve{k}\sigma}+
 \gamma^{B}_{\alpha\beta}(-\ve{k})\op{b}^{\dag(\alpha)}_{-\ve{k}-\sigma}
 \op{b}^{(\beta)}_{-\ve{k}-\sigma}
  - \sigma \gamma^{A}_{\alpha\beta}(\ve{k})\op{b}^{\dag(\alpha)}_{\ve{k}\sigma}
 \op{b}^{\dag(\beta)}_{-\ve{k}-\sigma}
 -\sigma  \bar{\gamma}^{A}_{\alpha\beta}(\ve{k})\op{b}^{(\alpha)}_{\ve{k}\sigma}
 \op{b}^{(\beta)}_{-\ve{k}-\sigma}
 \right\}\\
 &&-\frac{N_{c}}{2}\sum_{\ve{R}\alpha\beta}J_{\alpha\beta}(\ve{R})\left[|B_{\alpha\beta}(\ve{R})|^{2}-
 |A_{\alpha\beta}(\ve{R})|^{2}
 \right],
 \ea
\end{widetext}
\normalsize
where
\small
 \ba
 \!\!\gamma^{B}_{\alpha \beta}(\ve{k})\!\!&=&\!\!\frac12 \sum_{\ve{R}}\!
 J_{\alpha\beta}(\ve{R}) B_{\alpha\beta}(\ve{R}) e^{-i \ve{k}\cdot
 (\ve{R}+\ve{r}_{\alpha}-\ve{r}_{\beta})}\\
 \!\!\gamma^{A}_{\alpha \beta}(\ve{k})\!\!&=&\!\!\frac12 \sum_{\ve{R}}\!
 J_{\alpha\beta}(\ve{R}) A_{\alpha\beta}(\ve{R}) e^{-i \ve{k}\cdot
 (\ve{R}+\ve{r}_{\alpha}-\ve{r}_{\beta})}\\
 \! \!\bar{\gamma}^{A}_{\alpha \beta}(\ve{k})\!\!&=&\!\!\frac12 \sum_{\ve{R}}\!
 J_{\alpha\beta}(\ve{R}) \bar{A}_{\alpha\beta}(\ve{R}) e^{-i \ve{k}\cdot
 (\ve{R}+\ve{r}_{\alpha}-\ve{r}_{\beta})}.
 \ea
\normalsize

Now, we impose the constraint (\ref{eq:sb constraint}) in average over each
sublattice $\alpha$ by means of  Lagrange multipliers
$\lambda^{(\alpha)}$
 \ba
 \op{H}_{MF}\rightarrow
 \op{H}_{MF}+\op{H}_{\lambda}
 \ea
with
 \ba
 \op{H}_{\lambda}=\sum_{\ve{x}\alpha}\lambda^{(\alpha)}
 \left(\sum_{\sigma}\op{b}^{\dag(\alpha)}_{\ve{x}\sigma}\op{b}^{(\alpha)}_{\ve{x}\sigma}-2S\right).
 \ea

Using the symmetries (\ref{eq:simetrias1}) we can see that both kinds of
bosons ($\uparrow,\downarrow$)  give the same contribution to the Hamiltonian.
Then, we can perform the sum over  $\sigma$ to obtain
\begin{widetext}
 \ba
\nonumber \op{H}_{MF}&=&\frac12 \sum_{\ve{k}\alpha \beta}\left\{ (
\gamma^{B}_{\alpha \beta}(\ve{k})+\lambda^{(\alpha)}\delta_{\alpha
\beta})
\op{b}^{\dag(\alpha)}_{\ve{k}\uparrow}\op{b}^{(\beta)}_{\ve{k}\uparrow}+
 ( \gamma^{B}_{\alpha \beta}(-\ve{k})+\lambda^{(\alpha)}\delta_{\alpha \beta})
\op{b}^{\dag(\alpha)}_{-\ve{k}\downarrow}\op{b}^{(\beta)}_{-\ve{k}\downarrow}
 -
 \sigma \left(\gamma^{A}_{\alpha \beta}(\ve{k})\op{b}^{\dag(\alpha)}_{\ve{k}\uparrow}\op{b}^{\dag(\beta)}_{-\ve{k}\downarrow}
+  \bar{\gamma}^{A}_{\alpha
\beta}(\ve{k})\op{b}^{(\alpha)}_{\ve{k}\uparrow}\op{b}^{(\beta)}_{-\ve{k}\downarrow}
\right)
\right\}\\
\nonumber
&&-\frac{N_c}{2}\sum_{\ve{R}\alpha\beta}J_{\alpha\beta}(\ve{R})\left[|B_{\alpha\beta}(\ve{R})|^{2}-
 |A_{\alpha\beta}(\ve{R})|^{2}
 \right]-2SN_{c}\sum_{\alpha}\lambda^{(\alpha)}.
 \ea
\end{widetext}

It is convenient to define a vector operator
$\op{\ve{b}}^{\dag}(\ve{k})=\left(
\hat{\ve{b}}^{\dag}_{\ve{k}\uparrow},\hat{\ve{b}}_{-\ve{k}\downarrow}
\right)$ where
 \ba
  \hat{\ve{b}}^{\dag}_{\ve{k}\uparrow}&=& (\op{b}^{\dag(\alpha_{1})}_{\ve{k}\uparrow},\op{b}^{\dag(\alpha_{2})}_{\ve{k}\uparrow},...,
\op{b}^{\dag(\alpha_{n_{c}})}_{\ve{k}\uparrow})\\
\hat{\ve{b}}_{-\ve{k}\downarrow}&=&
(\op{b}^{\dag(\alpha_{1})}_{-\ve{k}\downarrow},\op{b}^{\dag(\alpha_{2})}_{-\ve{k}\downarrow},...,
\op{b}^{\dag(\alpha_{n_{c}})}_{-\ve{k}\downarrow})
 \ea
 and  $n_{c}$ is the number of atoms in the unit cell.
Now, we can write the Hamiltonian as
\ba
 \label{eq:HMF_compacto}
 H_{MF}&=&\sum_{\vk}\; \op{\ve{b}}^{\dag}(\ve{k})\cdot D(\vk) \cdot
\op{\ve{b}}(\ve{k})\\\nonumber
&-&(2S+1)N_{c}\sum_{\alpha}\lambda^{(\alpha)}- \langle H_{MF} \rangle,
\ea
where the   $2\, n_c \times 2\, n_c $ dynamical matrix $D(\ve{k})$  is given by
\small
 \ba
\nonumber
 D(\ve{k})\!=\!\left(
  \begin{tabular}{cc}
  $\gamma^{B}_{\alpha \beta}(\ve{k})+\lambda^{(\alpha)}\delta_{\alpha \beta}$ & $-\gamma^{A}_{\alpha \beta}(\ve{k})$ \\
$\gamma^{A}_{\alpha \beta}(\ve{k})$ & $\gamma^{B}_{\alpha
\beta}(\ve{k})+\lambda^{(\alpha)}\delta_{\alpha \beta}$
  \end{tabular}
\right).
 \ea
\normalsize
%
%
%

To diagonalize the Hamiltonian (\ref{eq:HMF_compacto}) we need to perform a para-unitary transformation of the
matrix  $D(\ve{k})$ that preserves the bosonic commutation relations. We can diagonalize the
Hamiltonian by defining the new operators
$\op{\ve{a}}=F \cdot \op{\ve{b}}$, where the matrix $F$ satisfy
\be
(F^{\dag})^{-1}\cdot \tau_3 \cdot (F)^{-1}= \tau_3 ,\quad \tau_3=
\left(
\begin{array}{cc}
I_{2\times 2} & 0\\
0 & -I_{2\times 2}
\end{array}
\right).
\ee
With this transformation, the Hamiltonian reads
\ba
\op{H}_{MF}=\sum_{\ve{k}} \op{\ve{a}}^{\dag}_{\ve{k}} \cdot \ve{E}(\ve{k}) \cdot \op{\ve{a}}_{\ve{k}}- 2(S+1)N_{c}\sum_{\alpha}\lambda^{(\alpha)}-\langle \op{H}_{MF} \rangle, \nonumber\\
\ea
where
\ba
\ve{E}(\ve{k})=\mbox{diag}(\omega_{1}(\ve{k}),...,\omega_{n}(\ve{k}),\omega_{1}(\ve{k}),
...,\omega_{n}(\ve{k})).
\ea
\begin{figure}[t]
\includegraphics[width=0.45\textwidth]{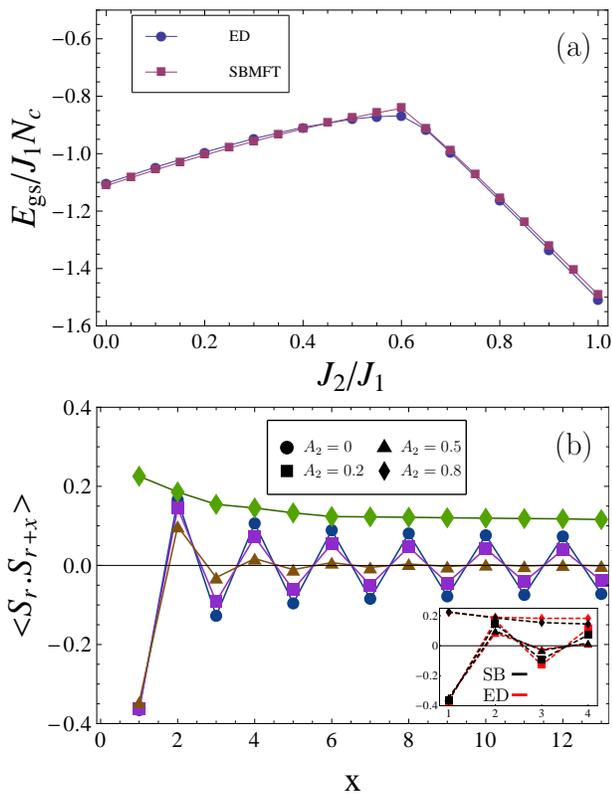}
\caption{(a): GS energy per unit cell for $S=\frac{1}{2}$ as a function of $J_{2}/J_{1}$ for a lattice
of 32 sites. The circles are exact results (ED) and the squares are
the SBMFT results. (b): Spin-Spin correlation function (SSCF) vs distance ${\bf X}$ in the {\it
zig-zag} direction obtained within SBMFT for a $32\times 32$ system in \cite{nos}.
For $0<J_2/J_1<0.41$, the SSCF correspond to the N\'eel phase with long-rage-order (LRO),
for $0.41< J_2/J_1<0.6$ the correlations are short ranged indicating a gap zone with sort-range-order (SRO),
and for $0.6< J_2/J_1$ the correlations correspond to the collinear phase
(ferromagnetic correlations in the zig-zag direction).
The inset  in Fig. (b) shows  the finite size results for the SSCF obtained by ED and SBMFT for 32 sites.
Lines are guides to the eye and different colors are used for clarity.}
\label{fig:comparacion_E}
\end{figure}
%

In term of the original bosonic operators, the mean field parameters are
\ba
\nonumber
A_{\alpha \beta}(\ve{R})\!\! &=&\!\! \frac{1}{2 N_{c}}\sum_{\ve{k}}
\left\{
e^{i\ve{k}(\ve{R}+\ve{r}_{\alpha}-\ve{r}_{\beta})} \langle
  \op{b}^{(\alpha)}_{\ve{k}\uparrow} \op{b}^{(\beta)}_{-\ve{k}\downarrow}\rangle \right.\\
\label{eq:sc1}
&-&\left. e^{-i\ve{k}(\ve{R}+\ve{r}_{\alpha}-\ve{r}_{\beta})} \langle  \op{b}^{(\alpha)}_{-\ve{k}\downarrow} \op{b}^{(\beta)}_{\ve{k}\uparrow}\rangle
 \right\} \\\nonumber
B_{\alpha \beta}(\ve{R})\!\! &=&\!\! \frac{1}{2 N_{c}}\sum_{\ve{k}}
\left\{
e^{i\ve{k}(\ve{R}+\ve{r}_{\alpha}-\ve{r}_{\beta})}
 \langle  \op{b}^{\dag(\beta)}_{\ve{k}\uparrow} \op{b}^{(\alpha)}_{\ve{k}\uparrow}\rangle  \right.\\
\label{eq:sc2}
&-&\left. e^{-i\ve{k}(\ve{R}+\ve{r}_{\alpha}-\ve{r}_{\beta})} \langle  \op{b}^{\dag (\beta)}_{-\ve{k}\downarrow} \op{b}^{(\alpha)}_{-\ve{k}\downarrow}\rangle
 \right\}
\ea
and the constraint in the number of bosons can be written in the momentum space as
%
\small
\ba
\label{eq:const1}
\sum_{\ve{k}} \left\{
 \langle  \op{b}^{\dag(\alpha)}_{\ve{k}\uparrow} \op{b}^{(\alpha)}_{\ve{k}\uparrow} \rangle +
\langle  \op{b}^{\dag(\alpha)}_{-\ve{k}\downarrow} \op{b}^{(\alpha)}_{-\ve{k}\downarrow} \rangle
 \right\}=2S N_{c},
\ea
\normalsize
where $N_c$ is the total number of unit cells and $S$ is the spin strength.
The mean field equations (\ref{eq:sc1}) and (\ref{eq:sc2}) must be solved in a self-consistent way together with the
constraints (\ref{eq:const1}) on the number of bosons.
Finding numerical solutions involves finding the roots of 24 coupled
nonlinear equations for the parameters $A$ and $B$, plus the additional constraints to determine
the values of the Lagrange multipliers $\lambda^{(\alpha)}$. We perform the calculations for
finite but very large lattices and  finally we extrapolate the results to the thermodynamic limit. We solve  numerically
 for several values of the frustration parameter $J_2/J_1$ and with the values obtained for the MF parameters and the Lagrange multipliers we
compute the energy and the new values for the MF parameters.
We repeat this self-consistent procedure until the energy and the MF parameters converge.
After reaching convergence we can compute all physical quantities like the energy, the spin-spin correlations and the excitation gap.
In order to support the analytical results of the MF approach, we have also performed exact diagonalization on finite systems with 18, 24
and 32 spins with periodic boundary conditions for $S=1/2$ using Spinpack \cite{spinpack}.

\section{Results}

In Fig.~\ref{fig:comparacion_E}(a) we show the ground state energy per unit cell
as a  function of the frustration for a system of 32 sites calculated by means
of SBMFT and ED, showing an excellent agreement between both approaches. The advantage
of the SBMFT is that it allows to study much larger systems:
we have studied different system sizes up to 3200 sites and extrapolated
to the thermodynamic limit.

\begin{figure}[t]
\includegraphics[width=0.45\textwidth]{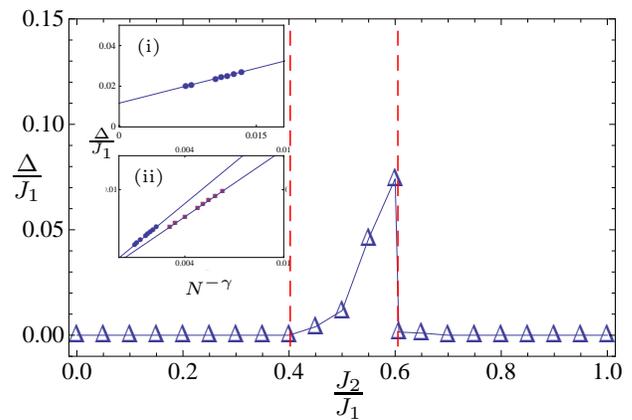}
   \caption{Gap in the boson dispersion as a function of $J_{2}/J_{1}$
for $S=1/2$ from Ref.\ \cite{nos}. In the region $J_{2}/J_{1}\sim 0.6$ the system remains gapped. Inset: finite size
scaling for the gap. (i) $J_2/J_1=0.5$ ($\gamma=0.6451$), (ii): Circles correspond to
$J_2/J_1=0.05$ ($\gamma=0.911$) and squares correspond to $J_2/J_1=0.35$ ($\gamma=0.758$).
}
\label{fig:gap800}
\end{figure}

For the present model we only find commensurate collinear phases
and for these phases, the wave vector $\ve{Q}_0={\bf Q}/2$ where the dispersion
relation has a minimum remains pinned at a commensurate point in the Brillouin
zone, independently of the value of the frustration $J_2/J_1$.
In the thermodynamic limit, a state with LRO is
characterized in the Schwinger boson approach by a condensation of bosons at
the wave vector ${\bf Q}_0$. This implies that the
dispersion of the bosons in a state with LRO is gapless. As we discussed
earlier, we solve the mean field equations for finite systems, then to detect LRO we
calculate the gap in the bosonic spectrum as a function of $J_2/J_1$  for
different system sizes and perform a finite size scaling finding a finite region
where the system remains gapped.

\begin{figure}[t]
\includegraphics[width=0.47\textwidth]{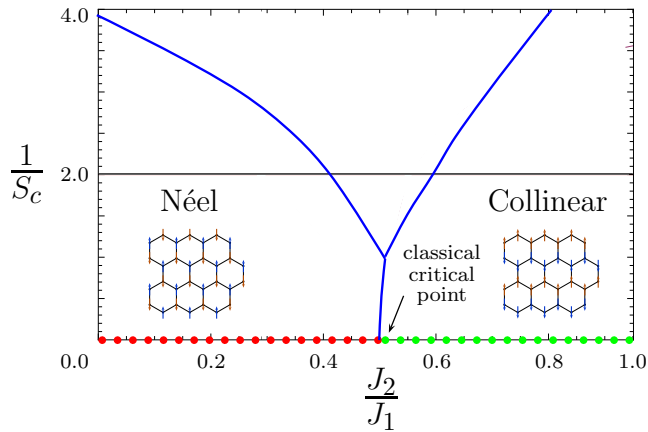}
\caption{(Color online) Inverse of the critical spin $S_c$ as a function of $J_2/J_1$ obtained using SBMFT by
us in \cite{nos}.
 For the $S=1/2$ case, there is a range $0.41< J_2/J_1<0.6$ where the system has a spin-gap indicating
a quantum disordered phase (see Fig.~\ref{fig:gap800}).
The dotted-line correspond to the classical limit $S\to\infty$ where the ground state correspond to
the Neel phase with $\ve{Q} =(0,0)$ and $\phi_{A}-\phi_{B}=\pi$ in the region $J_2/J_1<0.5$, while for
 $J_2/J_1>0.5$ the ground state correspond to the CAF phase characterized by
$\ve{Q}=(2\pi/\sqrt{3},0)$  and $\phi_{A}-\phi_{B}=\pi$.}
\label{fig:Sc_NAF}
\end{figure}
\begin{figure}[t!]
\includegraphics[width=0.45\textwidth]{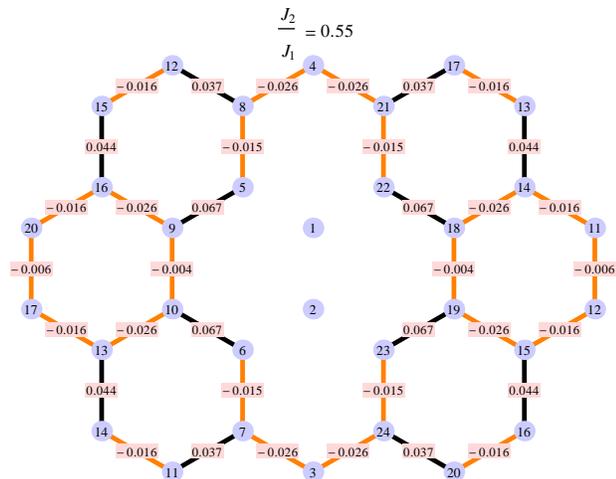}
   \caption{Dimer-dimer correlations $D_{(1,2),(k,l)}$
between the reference bond $\left(1,2\right)$ and bonds $\left(k,l\right)$
in the ground-state of the $N=24$ sample for $J_2/J_1=0.55$.
The number on bond $\left(k,l\right)$ indicates the value of $D_{(1,2),(k,l)}$
truncated to the two first significant digits. Full black (orange)
 lines  indicate positive (negative) values of $D_{(1,2),(k,l)}$.}
\label{fig:dimeros}
\end{figure}
The structure of the different phases can be understood calculating the
spin-spin correlation function (SSCF).\\
For $J_2/J_1<0.41$ the SSCF is antiferromagnetic in all directions and is long-ranged while for $0.6<J_2/J_1$
we have found ferromagnetic LRO correlations in the zig-zag direction that correspond to the
CAF phase. The most interesting result is in the intermediate region $0.41<J_2/J_1<0.6$ where
the results for the SSCF predict a quantum disordered state with a gap  in the bosonic dispersion and
the spin-spin correlation function shows N\'eel short range order.
A plot of the SSCF for $J_2/J_1=0,0.2,0.5$ and $0.8$ obtained within SBMFT is presented
in Fig.~\ref{fig:comparacion_E}b. In Fig.~\ref{fig:Sc_NAF} we show the ground state phase diagram
as a function of $1/S$ \cite{nos}. The classical phase diagram reduces to that shown in the line $1/S_{c} = 0$
of Fig.~\ref{fig:Sc_NAF} where two collinear phases meet at the classical critical point $J_2/J_1 =0.5$.
On the one hand, for $1/S$ smaller than a critical value $1/Sc(J_2/J_1)$, the correlation functions have
LRO, characterized by a condensation of bosons at the wave vector ${\bf Q}_0$.
On the other hand, when  $1/S$ is greater than $1/Sc(J_2/J_1)$, the correlation functions have SRO
indicating quantum disorder.\\
In the intermediate region  the results found with SBMFT predict a quantum disordered region
$0.41< J_{2}/J_{1}<0.6$. In this region a gap opens in the bosonic dispersion and the spin-spin
correlation function
shows N\'eel short range order followed by the LRO CAF phase for $J_2/J_1>0.6$.
In Fig.~\ref{fig:gap800}  we show the extrapolation of the boson gap as a function of the frustration.
The inset shows an example of the finite size scaling for different values of the frustration.

Previous results show that for $0.41< J_2/J_1<0.6$ the ground state has no magnetic order \cite{nos}.
The main question now is: Is this non  magnetic quantum phase a quantum  disordered one?
Or does it exhibit any other kind of non-magnetic order?.
To answer this question the knowledge of the spin-spin correlation function is not enough and one has to
check for other types of (non-magnetic) ordering patterns.

One kind of non magnetic order that can set in is the {\it dimer long-range order}.
The dimer operator on a pair of sites $\left(i,j\right)$ is defined as $\hat{\bf d}_{i,j}=1/4-\hat{\bf S}_i\cdot\hat{\bf S}_j$,
and one usually defines the dimer-dimer correlation between bonds $(i,j)$ and  $(k,l)$ as $D_{(i,j),(k,l)}=<\hat{\bf d}_{i,j}\hat{\bf d}_{k,l}>-<\hat{\bf d}_{i,j}><\hat{\bf d}_{k,l}>$.
In order to understand the nature of the ground state in the intermediate region, we have calculated de dimer-dimer correlation function defined above by means of exact diagonalization on a 24 sites cluster with periodic boundary conditions for $S=1/2$. The correlation pattern for dimers on first neighbor bonds is displayed in  Fig.~\ref{fig:dimeros}.
We can see that the exact dimer-dimer correlations show a rather fast decay suggesting that there is no dimer order in the groud state, though due to the small size of the cluster studied, this is not conclusive and we cannot discard other ordering patterns.

In summary, the results presented here further support
the existence of a region in the intermediate frustration regime where the system does not show quantum magnetic order for $S=1/2$.

{\it Note added}:  When this manuscript was completed we
became aware of two independent works providing an analysis
of the model  using a combination of exact diagonalizations \cite{Farnell,Capponi} and an effective quantum dimer model, as well as a self-consistent
cluster mean-field theory \cite{Capponi}. Several similar findings show
a good correspondence of both approaches.

Acknowledgements: This work was partially supported by the ESF grant INSTANS,
PICT ANPCYT (Grant No 2008-1426) and PIP CONICET (Grant No 1691).


\end{document}